\documentclass[a4paper,twocolumn]{revtex4}
\usepackage[utf8]{inputenc}
\usepackage{siunitx}
\usepackage{graphicx}
\usepackage{textcomp}
\usepackage[left=1.5cm, right=1.5cm, top=1.5cm, bottom=1.5cm]{geometry}
\usepackage{hyperref}
\hypersetup{colorlinks=true,linkcolor=blue,citecolor=blue}

\begin{document}

\title{Ultra-sensitive multi-species spectroscopic breath analysis \\for real-time health monitoring and diagnostics}

\author{Qizhong Liang$^{a,b}$, Ya-Chu Chan$^{a,c}$, P. Bryan Changala$^{d}$, David J. Nesbitt$^{a,b,c}$, Jun Ye$^{a,b,*}$, and Jutta Toscano$^{a,b,}$} 
\thanks{Corresponding authors: jutta.toscano@jila.colorado.edu, ye@jila.colorado.edu}
\affiliation{$^a$JILA, National Institute of Standards and Technology and University of Colorado, Boulder, CO 80309, USA}
\affiliation{$^b$Department of Physics, University of Colorado, Boulder, CO 80309, USA}
\affiliation{$^c$Department of Chemistry, University of Colorado, Boulder, CO 80309, USA}
\affiliation{$^d$Center for Astrophysics $|$ Harvard \& Smithsonian, Cambridge, MA 02138, USA}

\begin{abstract}
Breath analysis enables rapid, non-invasive diagnostics, as well as long-term monitoring, of human health through the identification and quantification of exhaled biomarkers. Here, for the first time, we demonstrate the remarkable capabilities of mid-infrared (mid-IR) cavity-enhanced direct frequency comb spectroscopy (CE-DFCS) applied to breath analysis. We simultaneously detect and monitor as a function of time four breath biomarkers --- CH$_3$OH, CH$_4$, H$_2$O and HDO --- as well as illustrating the feasibility of detecting at least six more (H$_2$CO, C$_2$H$_6$, OCS, C$_2$H$_4$, CS$_2$ and NH$_3$) without modifications to the experimental apparatus. We achieve ultra-high detection sensitivity at the parts-per-trillion level. This is made possible by the combination of the broadband spectral coverage of a frequency comb, the high spectral resolution afforded by the individual comb teeth, and the sensitivity enhancement resulting from a high-finesse cavity. Exploiting recent advances in frequency comb, optical coating, and photodetector technologies, we can access a large variety of biomarkers with strong carbon-hydrogen bond spectral signatures in the mid-IR. 
\end{abstract}

\maketitle

\section*{Significance statement}
Determining the identity and concentration of the molecules present in breath is a powerful tool to assess the overall health of a person, analogously to the use of blood tests in clinical medicine but in a faster and less invasive manner. The presence of a particular molecule (or combination of molecules) can indicate the presence of a certain health condition or infection, facilitating a diagnosis. Monitoring the concentration of the molecules of interest over time can help track the development (or recurrence) of a condition, as well as the effectiveness of the administered treatment. In order to make breath analysis more accessible and widely adopted, we explore a new technique to simultaneously measure several molecules in breath with exceptional sensitivity and specificity.

\section{Introduction}
Breath analysis is an exceptionally promising and rapidly developing field of research which examines the molecular composition of exhaled breath \cite{Henderson2018, Lourenco2014, Risby2006, Modak2011, Das2020, Wang2009}. The hundreds of different gases that are present in exhaled breath include inorganic compounds as well as volatile organic compounds (VOCs) and can either result from internal metabolic activity (endogenous emissions) or external factors such as food consumption or environmental exposure (exogenous emissions). Despite its distinctive advantages of being a rapid, non-invasive technique and its long history dating back to Hippocrates, breath analysis has not yet been as widely deployed for routine diagnostics and monitoring as other methods, such as blood-based analysis. This is partly due to the experimental challenges of dealing with extremely small amounts of gas-phase molecules --- in the parts-per-million (ppm) to parts-per-billion (ppb) range for most VOCs --- and partly due to the relative scarcity of large-scale clinical studies that can reliably correlate specific diseases with biomarkers present in breath. Nevertheless, through close collaborations between instrument developers, breath analysis experts, and clinicians, the field of breath analysis is fast-approaching its goal of enabling real-time, non-invasive early detection and long-term monitoring of temporary and permanent health conditions \cite{Henderson2018, Risby2006}. Several biomarkers present in breath have been associated with specific conditions --- for instance nitrogen monoxide with asthma, acetone with diabetes, and ammonia with renal failure \cite{Das2020} --- and breath is increasingly being used to track diseases and infections, both bacterial and viral \cite{Jones2020}. Recently, three studies have demonstrated the use of breath analysis to discriminate between SARS-CoV-2 infected patients and patients affected by other conditions (including asthma, chronic obstructive pulmonary disease, bacterial pneumonia, and cardiac conditions) \cite{Ruszkiewicz2020,Shan2020} or Influenza-A infected patients \cite{Steppert2020}. The possibility of real-time testing for highly infectious diseases in a non-invasive manner, without the need for chemical reagents and complex laboratory facilities, is particularly appealing in view of the current global pandemic.  

Technologies being explored and adopted for breath analysis include mass spectrometry, nanomaterial-based sensors and laser spectroscopy. To date, the most widely used analytical technique in breath research is gas chromatography combined with mass spectrometry (GC-MS), which allows for the sensitive detection of hundreds of exhaled molecules, albeit with relatively long analysis times (tens of minutes) limited by the elution time of the various species. On the other hand, selected ion flow-tube mass spectrometry (SIFT-MS) and proton-transfer reaction mass spectrometry (PTR-MS) allow for real-time breath analysis at the expense of a reduced number of simultaneously detectable molecules \cite{Smith2014}. Sensor arrays offer an inexpensive and practical alternative for identifying the presence of a class of compounds based on their functional groups, but they generally do not permit identification of the specific molecules present in the samples \cite{Broza2013,Shan2020}. Laser spectroscopy is intrinsically fast ($<<$ second timescale), allowing breath-cycle-resolved (i.e., respiratory-phase-resolved) sampling of breath with high precision and absolute accuracy. Achieving high sensitivity requires both signal enhancement and noise reduction: the former is attained using multi-pass cells or high-finesse cavities, whilst the latter is accomplished through intensity or frequency modulation techniques. Among others, tunable diode laser absorption spectroscopy, cavity ring-down spectroscopy, cavity-enhanced absorption spectroscopy, and photoacoustic spectroscopy have all successfully been employed in breath analysis but are typically limited in tunability and therefore in the number of detectable analytes \cite{Henderson2018}. Cavity-enhanced direct frequency comb spectroscopy (CE-DFCS) offers substantially enhanced capabilities for the simultaneous detection of multiple species due to the combination of high spectral resolution, wide spectral coverage, and high sensitivity \cite{Adler2010,Maslowski2014,Changala2018,Weichman2019,Thorpe2008a,Foltynowicz2011}. An early study from 2008 demonstrated this by detecting carbon monoxide, carbon dioxide, methane, ammonia, and water in breath samples by CE-DFCS \cite{Thorpe2008}. This previous work measured vibrational (mainly first overtone) transitions in the near-infrared region of the spectrum, from \SI{1.5}{\micro \meter} to \SI{1.7}{\micro \meter}. 

Here, we report a two orders of magnitude improvement in the detection sensitivity for multiple species relevant to breath analysis by using CE-DFCS in the mid-IR molecular fingerprint region. Exploiting recent advances in frequency comb, high-reflectivity optical coating, and photodetector technologies, we can detect a large variety of biomarkers simultaneously, sensitively, and unambiguously, providing exciting prospects to connect breath to a range of biological functions and diseases.

\section{Results}\label{Results}

A frequency comb is a coherent light source consisting of hundreds of thousands of discrete, equidistant, and narrow frequency modes, often referred to as comb teeth. Despite the large number of comb teeth spanning a broad wavelength range, the exact frequency of each individual line can be precisely controlled through frequency stabilization \cite{Ye2004,Schliesser2012,Fortier2019,Diddams2020}. The simultaneous highly-resolved and broadband nature of frequency combs make them ideal light sources for spectroscopy. Additionally, the high degree of control over the exact frequency of each comb tooth enables efficient coupling to a high-finesse enhancement cavity by matching the frequencies of the comb teeth with those of the resonant cavity modes. The combination of frequency combs and high-finesse cavities adds high sensitivity to the previously mentioned benefits of high resolution and broadband spectral coverage. To exploit all three of these features for breath analysis applications, we employ a high power mid-IR frequency comb \cite{Adler2009} in combination with an ultra-high-finesse mid-IR enhancement cavity and near-shot-noise-limited detection to achieve exceptionally high sensitivity for a range of molecules present in breath.

\begin{figure}[t!]
\centering
\includegraphics[width=8.65cm]{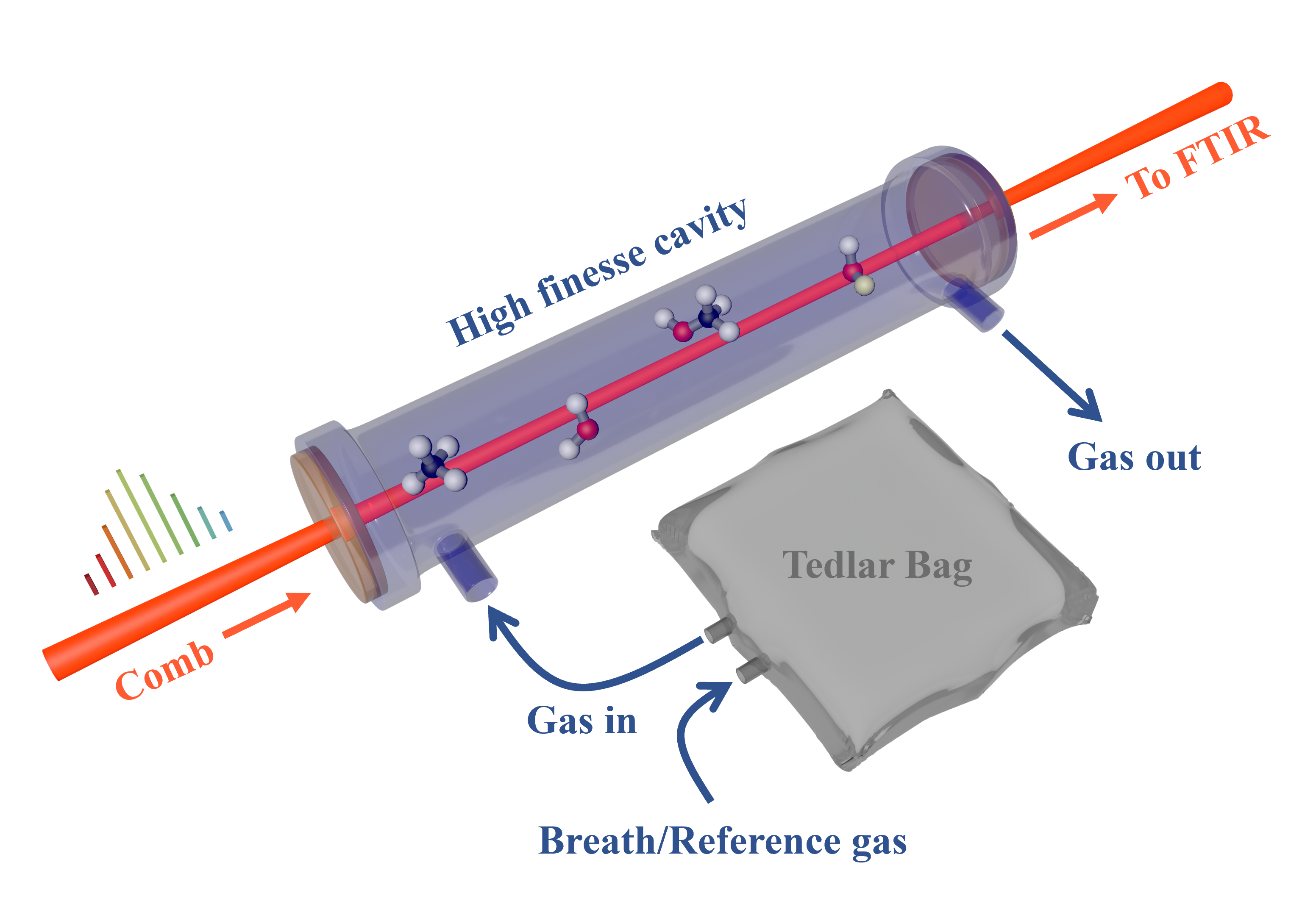}
\caption{Schematic representation of the experimental apparatus. Breath --- or, alternatively, a reference gas --- is collected into a Tedlar bag and subsequently loaded into a high-finesse cavity comprised of a pair of high reflectivity mirrors. A mid-infrared frequency comb is resonantly coupled into the cavity and interacts with the molecules present inside the cavity during several thousand round trips. The cavity transmitted comb light is analyzed by a Fourier-transform infrared (FTIR) spectrometer to determine the identity and concentration of the molecular species present in the gas sample.}
\label{fig:ExpApp}
\end{figure}

\begin{figure*}[th]
\centering
\includegraphics[width=17.75cm]{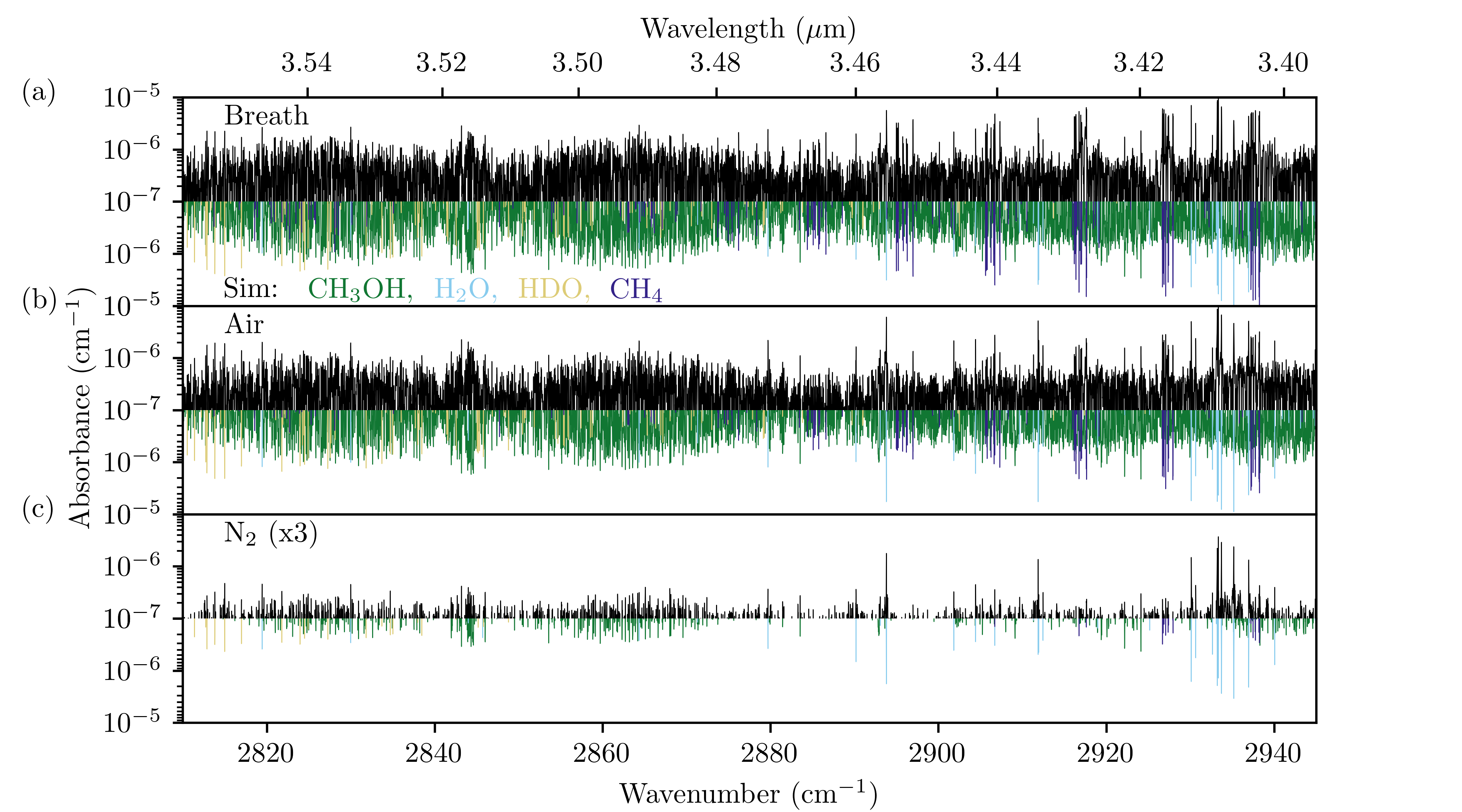}
\caption{Experimental and simulated absorbance spectra of breath (a), lab-environment air (b) and nitrogen gas (c). The experimental data is depicted in black while the simulated data is shown inverted with the colour corresponding to the absorbing species: methanol (green), water (light blue), partially-deuterated water (gold) and methane (dark blue). Note that the molecules detected in the nitrogen gas spectrum (which has been magnified by a factor of 3) are remaining impurities following the cleaning protocol performed between measurements (see text).}
\label{fig:FullScans}
\end{figure*}

The breath sample or a reference gas is introduced directly into the enhancement cavity where it absorbs part of the comb light; the absorbed frequency components are subsequently detected using a Fourier-transform spectrometer (Fig. \ref{fig:ExpApp}). Typical absorption spectra obtained in this way are shown in Fig. \ref{fig:FullScans}. For the breath measurements (Fig. \ref{fig:FullScans}(a)), the volunteer (one of the co-authors) is asked to inhale to total lung capacity before blowing exclusively the second half of the exhalation (end-tidal breath sampling) into the reusable Tedlar bag through a mouthpiece. Measurements of the background laboratory air (Fig. \ref{fig:FullScans}(b)) are taken for comparison and constitute a baseline for the breath measurements. The nitrogen gas measurements (Fig. \ref{fig:FullScans}(c)), on the other hand, are taken before and after each breath or air measurement, following the cleaning of the Tedlar bag and of the enhancement cavity by flowing nitrogen gas through the system for 5 minutes (with a flow of $\approx~$\SI{30}{stdl/min}). These serve to check the cleanliness of the system and control against cross-contamination from one sample to the next. 

Within the examined spectral window (\SIrange{2810}{2945}{cm^{-1}}), we observe absorption features from four molecular species: methanol (CH$_{3}$ $s $-stretch, $\nu_{3}$, with Q-branch at \SI{2844}{cm^{-1}} and CH$_{3}$ $d $-stretch, $\nu_{2}$ and $\nu_{9}$), methane (P-branch of the $\nu_{3}$ band), water (first overtone of the bend, 2$\nu_{2}$) and partially-deuturated water (first overtone of the bend, 2$\nu_{2}$). We fit the experimental absorption spectra to those calculated from the HITRAN database {\cite{Gordon2017,Kochanov2016}} to extract the concentration of each species. These simulated spectra are shown in Fig. \ref{fig:FullScans} (see SI for more details). The multi-line, global fitting procedure we adopt allows us to fit all species at once across the entire spectral window, enabling a precise determination of the molecular concentrations despite the presence of overlapping spectral features. We find the measured concentrations of methanol and methane in the volunteer's breath sample (CH$_3$OH~=~\SI{4.53(4)}{ppm}, CH$_4$~=~\SI{1.49(2)}{ppm}) to be higher than the respective background concentrations in air (CH$_3$OH~=~\SI{3.21(2)}{ppm}, CH$_4$~=~\SI{0.69(1)}{ppm}). In contrast, we measure similar concentrations of water and partially-deuturated water in breath (H$_2$O~=~\SI{1.76(3)}{}\textperthousand, HDO~=~\SI{1.83(8)}{ppm}) and air samples (H$_2$O~=~\SI{2.09(2)}{}\textperthousand, HDO~=~\SI{1.50(5)}{ppm}). The concentrations of the four species in the nitrogen gas sample are one or two orders of magnitude lower than those measured in the breath and air samples (CH$_3$OH~=~\SI{0.22(2)}{ppm}, CH$_4$~=~\SI{18(1)}{ppb}, H$_2$O~=~\SI{0.23(2)}{}\textperthousand, HDO~=~\SI{96(5)}{ppb}), demonstrating the effectiveness of the aforementioned cleaning protocol.  

\begin{table*}[th]
\centering
\footnotesize
\begin{tabular}{c|cccc|c|c}
\textbf{Molecule} & \multicolumn{4}{c|}{\textbf{Achieved}}                                           & \textbf{Potential}              & \textbf{Clinical relevance} \cite{Broza2013,Risby2006,Wang2009} \\
                                   & \multicolumn{4}{c|}{\textbf{(2810--2945 cm$^{-1}$)}}                             & \textbf{(2100--3600 cm$^{-1}$)} &                                              \\ \cline{2-6}
                                   & Max. absorption               & Single-elem.  & Multi-elem.   & Spectral         & Max. absorption                 &                                              \\
                                   & cross section                 & detection     & detection     & element          & cross section                   &                                              \\
                                   & (\SI{}{cm^{2}~molecule^{-1}}) & limit         & limit         & utilization (\%) & (\SI{}{cm^{2}~molecule^{-1}})   &                                              \\ \hline
H$_2$CO                            & \SI{4e-18}{}                  & \SI{280}{ppt}   & \SI{76}{ppt} & \SI{98.30}       & \SI{4e-18}{}                    & Lung cancer                                  \\
                                   & (at \SI{2815}{cm^{-1}})       &               &               &                  & (at \SI{2815}{cm^{-1}})         &                                              \\
                                   &                               &               &               &                  &                                 &                                              \\
C$_2$H$_6$                         & \SI{1e-18}{}                  & \SI{1.5}{ppb}   & \SI{230}{ppt} & \SI{99.08}       & \SI{1e-17}{}                    & Asthma                                       \\
                                   & (at \SI{2907}{cm^{-1}})       &               &               &                  & (at \SI{2983}{cm^{-1}})         & Chronic obstructive pulmonary disease        \\
                                   &                               &               &               &                  &                                 & Inflammatory bowel disease                   \\
CH$_4$*                            & \SI{4e-18}{}                  & \SI{1.4}{ppb}   & \SI{390}{ppt} & \SI{60.98}       & \SI{1e-17}{}                    & Intestinal problems                          \\
                                   & (at \SI{2938}{cm^{-1}})       &               &               &                  & (at \SI{3058}{cm^{-1}})         & Bacteria and colonic fermentation            \\
                                   &                               &               &               &                  &                                 &                                              \\
CH$_3$OH*                          & \SI{3e-19}{}                  & \SI{2.5}{ppb}  & \SI{440}{ppt} & \SI{99.33}       & \SI{5e-19}{}                    & Hangover                                     \\
                                   & (at \SI{2924}{cm^{-1}})       &               &               &                  & (at \SI{2982}{cm^{-1}})         &                                              \\
                                   &                               &               &               &                  &                                 &                                              \\
OCS                                & \SI{8e-19}{}                  & \SI{3.0}{ppb}   & \SI{550}{ppt} & \SI{50.11}       & \SI{8e-19}{}                    & Liver fetor                                  \\
                                   & (at \SI{2926}{cm^{-1}})       &               &               &                  & (at \SI{2926}{cm^{-1}})         &                                              \\
                                   &                               &               &               &                  &                                 &                                              \\
HDO*                               & \SI{8e-19}{}                  & \SI{2.9}{ppb}  & \SI{830}{ppt}   & \SI{44.60}        & \SI{2e-18}{}                    & Measurement of total body water weight       \\
                                   & (at \SI{2815}{cm^{-1}})       &               &               &                  & (at \SI{2720}{cm^{-1}})         & (through measurement of HDO/H$_2$O           \\
                                   &                               &               &               &                  &                                 & concentration ratio)                         \\
C$_2$H$_4$                         & \SI{9e-20}{}                  & \SI{44}{ppb}  & \SI{12}{ppb}  & \SI{11.58}        & \SI{6e-19}{}                    & Lipid peroxidation                           \\
                                   & (at \SI{2943}{cm^{-1}})       &               &               &                  & (at \SI{3085}{cm^{-1}})         &                                              \\
                                   &                               &               &               &                  &                                 &                                              \\
CS$_2$                             & \SI{2e-20}{}                  & \SI{39}{ppb} & \SI{10}{ppb}  & \SI{15.04}       & \SI{2e-18}{}                    & Halitosis                                    \\
                                   & (at \SI{2840}{cm^{-1}})       &               &               &                  & (at \SI{2178}{cm^{-1}})         & Schizophrenia                                \\
                                   &                               &               &               &                  &                                 &                                              \\
NH$_3$                             & \SI{1e-20}{}                  & \SI{120}{ppb}   & \SI{63}{ppb} & \SI{2.22}        & \SI{8e-19}{}                    & Asthma                                       \\
                                   & (at \SI{2893}{cm^{-1}})       &               &               &                  & (at \SI{3336}{cm^{-1}})         & Halitosis                                    \\
                                   &                               &               &               &                  &                                 & Chronic renal failure/uremia                 \\
H$_2$O*                            & \SI{3e-21}{}                  & \SI{930}{ppb}   & \SI{490}{ppb}   & \SI{0.52}        & \SI{2e-18}{}                    & Measurement of total body water weight       \\
                                   & (at \SI{2935}{cm^{-1}})       &               &               &                  & (at \SI{3589}{cm^{-1}})         & (through measurement of HDO/H$_2$O           \\
                                   &                               &               &               &                  &                                 & concentration ratio)                        
\end{tabular}
\caption{Molecules absorbing in the spectral region probed in this work that are relevant for breath analysis, with the asterisk (*) denoting molecules detected in this work. The maximum absorption cross section --- with the corresponding wavenumber shown in brackets underneath --- indicates the strongest absorption peak in the probed spectral range. The single-element detection limit is determined at the single frequency with highest signal-to-noise ratio within the probed spectral range. Using the transition lines contained in the entire spectrum --- instead of the absorption at a single frequency --- yields the multi-element detection limit, which is (depending on the molecule) between 2 and 7 times lower than the single-element detection limit. The spectral element utilization column reports the percentage of the entire spectrum that contains absorbing features for each molecule, using an arbitrary threshold cross section of \mbox{$1 \times 10^{-22}$ cm$^2$ molecule$^{-1}$}. With the spectral tunability of the mid-IR frequency comb covering the range \mbox{2100--3600 cm$^{-1}$}, the reported detection limits could be further improved by probing the molecules closer to their maximum absorption features, shown under the ``Potential'' column which takes into account a wider spectral region (\mbox{2100--3600 cm$^{-1}$}) compared to the one probed in this work (\mbox{2810--2945 cm$^{-1}$}). ppt: parts-per-trillion; ppb: parts-per-billion; ppm: parts-per-million; the average simulated Lorentzian and Gaussian linewidth components are 0.31(8)~GHz and 0.21(6)~GHz, respectively. (See SI for more details).}
\label{table:DetLim}
\end{table*}

Although the breath samples analysed in this work only contain the four species discussed above, several other molecules absorb within the probed spectral region and could therefore also be detected if they were present in the analysed samples (Table \ref{table:DetLim}). From the noise floor of our spectra, we estimate the absorption sensitivity per spectral element of our experiment to be \SI{5(1)e-11}{cm^{-1} Hz^{-1/2}} (see SI for more details). Such exceptionally high sensitivity corresponds to especially low minimum detectable concentrations, which are reported for several species of interest in Table \ref{table:DetLim}. In particular, detection sensitivity at the part-per-trillion level is achievable in our apparatus for six molecules of clinical relevance to breath analysis: H$_2$CO, C$_2$H$_6$, CH$_4$, CH$_3$OH, OCS and HDO. Parts-per-billion-level sensitivity is attainable for the other four molecules: C$_2$H$_4$, CS$_2$, NH$_3$ and H$_2$O. Given the high concentration of water vapor present in breath, its slightly higher detection limit (\SI{490}{ppb}) is not detrimental but rather advantageous, since it mitigates potential saturation issues without affecting the ability to detect its presence in breath samples. Compared to a single-frequency spectroscopy experiment (single-element detection limit in Table \ref{table:DetLim}), the simultaneous detection of a broad range of frequencies demonstrated in this study affords a multi-element detection limit which is between 2 and 7 times lower, depending on the molecule. In essence, CE-DFCS not only enables the detection of several molecules simultaneously, but it also lowers each of their minimum detectable concentrations. 

\begin{figure*}[th]
\centering
\includegraphics[width=17.75cm]{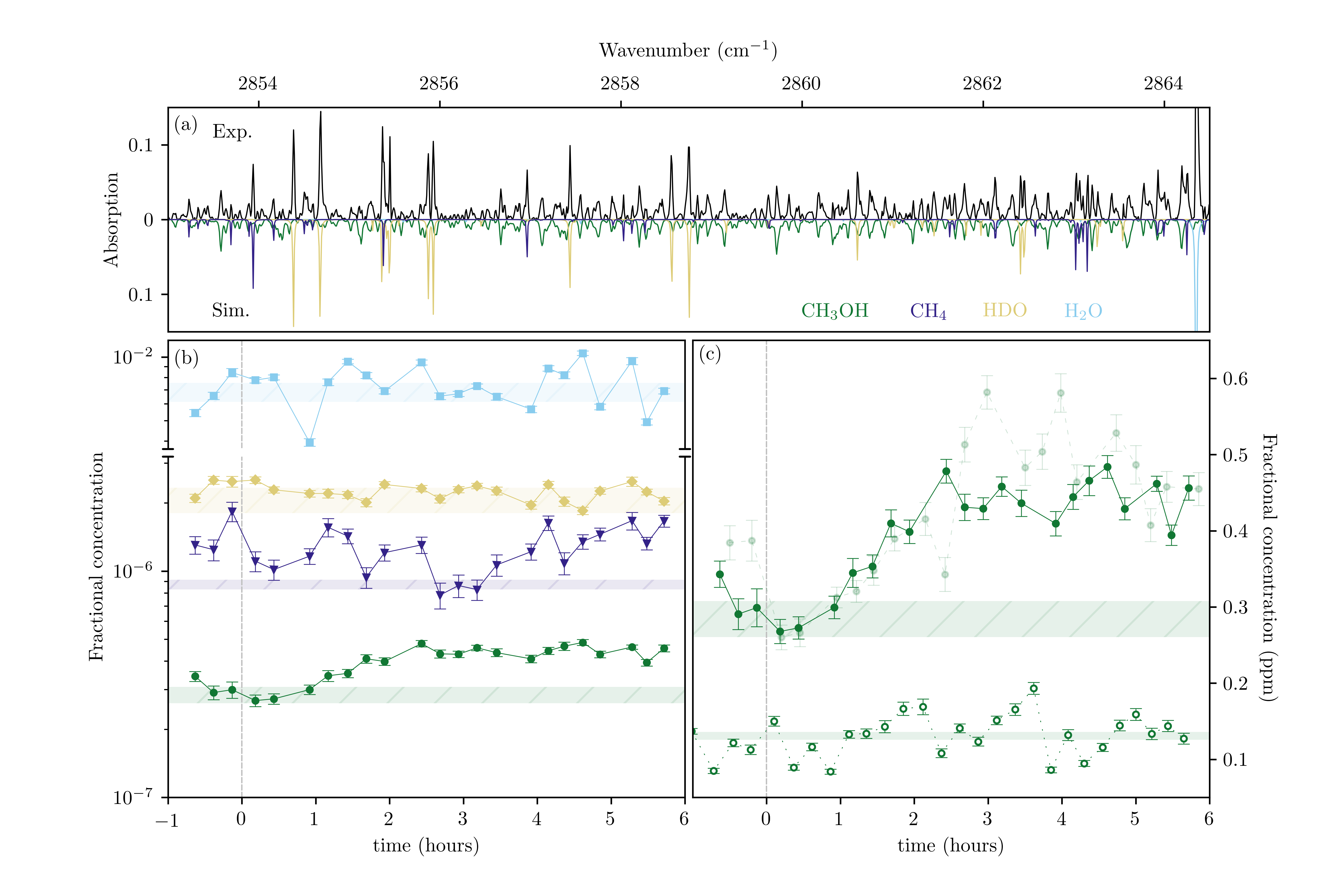}
\caption{Measurement of concentrations over time. Panel (a): Sample spectrum of breath in the spectral region used for this measurement. The experimental data is depicted in black while the simulated data is shown inverted with the colour corresponding to the absorbing species: methanol (green), water (light blue), partially-deuterated water (gold) and methane (dark blue). Panel (b): Fractional concentration of the four absorbing species as a function of time. The data points correspond to the concentrations obtained from the breath samples, with the error bars showing the 1$\sigma$ uncertainty from the fit to the experimental spectra. The shaded hatched regions represent the mean and standard error of the concentrations measured for the lab air samples (background measurements). Note the large dynamic range of measured fractional concentrations in the order of \SI{e5}{}. Panel (c): Methanol concentration as a function of time for breath (filled circles) and nitrogen gas (empty circles). The shaded hatched (non-hatched) regions represent the mean and standard error of the lab air (nitrogen gas) measurements. Note that the breath data points (filled circles) and shaded hatched region are the same as in panel (b) but plotted on a linear scale. For reference, a second set of data for methanol concentration in breath collected on a different day following the same protocol is also shown (pale filled circles); the mean and standard error of the background methanol fractional concentration for the second data set (\SI{3.0(3)e-7}{}) is comparable to the one shown. The vertical dashed line in panels (b) and (c) denotes the time of fruit consumption.}
\label{fig:FullFig}
\end{figure*}

To demonstrate the feasibility of applying CE-DFCS to real-time breath monitoring, we tracked the concentration of the molecules present in the breath of the same volunteer before and after the consumption of ripe fruit (specifically, 10 ripe baby bananas, $\approx~$\SI{0.5}{kg}). For the duration of the experiment (7 hours), as well as for a period of 12 hours before the beginning of the experiment, the volunteer did not consume any other food or drink. To achieve better time resolution, here we focus on a narrower spectral range (\SIrange{2853.0}{2864.5}{cm^{-1}}, Fig. \ref{fig:FullFig}(a)) whilst maintaining the ability to track all four molecules observed in Fig. \ref{fig:FullScans}. A breath spectrum is recorded approximately every 15 minutes and an air spectrum approximately every 1.5 hours, for a total of 23 breath and 5 air spectra collected. Between each measurement, the Tedlar bag and enhancement cavity are cleaned following the procedure outlined above and then a nitrogen gas spectrum is collected to check for remaining impurities (for a total of 28 N$_2$ spectra). Throughout the experiment, the breath fractional concentrations of water, partially-deuturated water and methane fluctuate around average values of \SI{7.4(3)}{}\textperthousand, \SI{2.24(4)}{ppm} and \SI{1.26(6)}{ppm}, respectively (Fig. \ref{fig:FullFig}(b)). However, the breath fractional concentration of methanol increases from \SI{311(13)}{ppb} to \SI{444(7)}{ppb} after the consumption of fruit, levelling off after approximately 2.5 hours (Fig. \ref{fig:FullFig}(b),(c)). A similar trend is observed on a second day of measurements following the same experimental protocol (Fig. \ref{fig:FullFig}(c)). The observed increase in methanol following fruit consumption is consistent with previous work and results from endogenous production of methanol \cite{Lagg1994, Taucher1995}. 

\section{Discussion}\label{Discussion}

The ultra-high molecular detection sensitivity achieved in this study stems from three factors: (1) the exceptionally high absorption sensitivity per spectral element of \SI{5(1)e-11}{cm^{-1} Hz^{-1/2}}, (2) the large absorption cross sections available in the mid-IR spectroscopic region (up to 4 orders of magnitude larger than those in the near-IR), and (3) the simultaneous measurement of multiple spectral frequencies enabled by the wide spectral coverage of a frequency comb. These factors are responsible for the 2 orders of magnitude improvement over the sensitivity per spectral element recorded by the previous CE-DFCS study of breath, which probed the near-infrared spectral region instead \cite{Thorpe2008}. Compared to previous direct frequency comb spectroscopy work in the mid-IR region, our detection limits for methane and methanol are 5 and 8 times lower, respectively, and our predicted detection limit for formaldehyde is 30 times lower \cite{Adler2010a}. In this case, the critical factor for such remarkable improvement comes from using a high-finesse enhancement cavity instead of a multipass cell. Overall, our sensitivity is comparable to the highest sensitivity per spectral element achieved in the mid-IR region by CE-DFCS to date (\SI{6.9e-11}{cm^{-1} Hz^{-1/2}} \cite{Foltynowicz2011a}). It is worth noting that, whilst the goal of this work is to enable the simultaneous detection of numerous different species, it is possible to achieve even lower detection limits than those reported here by probing the spectral region with the strongest absorption for each individual molecule (Table \ref{table:DetLim}). This could be easily attained with the current set-up, upon replacement of the high reflectivity cavity mirrors. In general, the examined spectral region can be selected depending on the specific clinical needs.  

Detecting multiple species simultaneously is particularly desirable in the exploratory phase of breath research, as well as with the view to monitor health through breath analysis as it is currently done through blood analysis. One of the main drawbacks of using optical methods for breath analysis is the limited number of molecular species that can be detected compared to mass spectrometric techniques. Indeed, all but one of the optical-based breath tests approved by the US Food and Drug Administration (FDA) detect a single species \cite{Henderson2018}. Here, we push this limit and demonstrate that CE-DFCS can potentially detect tens of species simultaneously, provided that they exhibit sharp absorption features (see Methods section). The latter requirement makes this technique most suitable for detecting light biomarkers, which --- depending on the exact ionization technique employed --- can be challenging to sensitively and reliably detect by mass spectrometry. This is due to the fragmentation of large molecules, with fragments potentially appearing at the same mass-to-charge ratio as smaller molecules in mass spectra, making it difficult to disentangle the contribution to the signal from light biomarkers and that from fragments of heavier biomarkers. Additionally, IR absorption spectroscopy is both isomer- and isotopologue-specific, whereas mass spectrometry is not typically able to distinguish between isomers. In these respects, frequency comb spectroscopy and mass spectrometry are complementary techniques for breath analysis. Finally, it is worth mentioning that it is possible to use independent laser sources to track different molecules in breath simultaneously, as it has been elegantly demonstrated for the real-time monitoring of O$_2$, CO$_2$ and H$_2$O to track oxygen consumption during an aortic aneurysm repair operation \cite{Ciaffoni2016} and of NO and CO$_2$ to record accurate NO expirograms \cite{Petralia2020}. Using a single broadband laser source to detect all molecules, such as it is shown in this work with a frequency comb, allows more accurate determination of the relative concentrations of the probed molecules, and may therefore be preferable for some applications. 

\section{Conclusion}\label{Conclusion}

By operating in the mid-IR molecular fingerprint region, we have demonstrated the unique and powerful capabilities of mid-infrared cavity-enhanced direct frequency comb spectroscopy (CE-DFCS) for breath analysis, including: 
\begin{itemize} \vspace{-2mm}
\item Ultra-high detection sensitivity at the parts-per-trillion (ppt) and parts-per-billion (ppb) levels;\vspace{-2.5mm}
\item Simultaneous detection of 4 breath biomarkers, with the potential to detect at least 6 more;\vspace{-2.5mm}
\item Real-time, simultaneous monitoring of biomarkers to track the physiological response to stimuli (food);\vspace{-2mm}
\end{itemize}
This is made possible by the combination of the broadband spectral coverage of a frequency comb, the high spectral resolution afforded by the individual comb teeth, and the sensitivity enhancement resulting from the high-finesse cavity. Probing the mid-IR spectral region is beneficial due to the presence of strong fundamental vibrational transitions, whilst probing a large spectral range allows us to detect different molecules simultaneously, as well as to improve our detection limit. The simultaneous detection of different species leads to faster measurements but also to more reliable determination of relative and absolute concentrations. In conclusion, we have shown that this technique offers unique advantages and opportunities for the detection of light biomarkers in breath and it is poised to facilitate real-time, non-invasive monitoring of breath for clinical studies as well as for early detection and long-term monitoring of temporary and permanent health conditions. 

\section{Methods}

\subsection*{Optical frequency comb}
The mid-infrared (mid-IR) frequency comb used in this study is generated from a singly-resonant optical parametric oscillator (OPO) synchronously pumped by a femtosecond Yb:fiber laser at a repetition rate of \SI{136}{MHz}. The center frequency of the mid-IR comb is tunable from \SI{2600}{cm^{-1}} to \SI{3100}{cm^{-1}} by adjusting the OPO crystal position inside the oscillator cavity. In this work, we have fixed the center frequency to be near \SI{2880}{cm^{-1}} with the intent of detecting a large number of molecular species of clinical relevance, instead of maximizing the detection sensitivity for any particular species to its ultimate limit. We typically use 50--\SI{100}{mW} of mid-IR optical power. More details on the light source can be found in Ref. \cite{Adler2009}. 

\subsection*{Cavity-enhanced spectroscopy}
The frequencies of the mid-IR comb and the enhancement cavity are fully stabilized and referenced to an ultra-low phase noise quartz oscillator, which is slaved to a cesium (Cs) clock (see SI for details on the servo loop implementations). The cavity free spectral range (FSR) is maintained at twice the comb repetition rate, which results in the filtering of every other cavity incident comb line and doubling of the repetition rate of the transmitted comb light to \SI{272}{MHz}. The cavity finesse ranges between 6000 and 8000 in the probed spectral region (see SI for more details), and the spectral coverage of the high reflectivity mirrors coating extends from \SI{2800}{cm^{-1}} to \SI{3300}{cm^{-1}}. The cavity transmitted comb light is detected by a home-built Fourier-transform IR spectrometer with an autobalancing HgCdZnTe photodetector, referenced to a passively stable \SI{1064}{nm} continuous-wave laser. By employing the properties of the known instrument function (a sinc function) and setting the spectral resolution to half the comb spacing (\SI{136}{MHz}), the intensity of each individual comb line can be extracted. The resulting comb-mode-resolved spectra have an effective spectral resolution limited by the linewidth of the individual comb teeth ($\approx~$\SI{50}{kHz}), rather than by the resolution of the spectrometer (\SI{4.5e-3}{cm^{-1}}, which is our data sampling frequency interval) \cite{Maslowski2016}. While the spectral bandwidth of the incident comb light covers more than \SI{100}{cm^{-1}}, we limit only a narrow spectral portion of $\approx~$\SI{10}{cm^{-1}} to be simultaneously coupled into the enhancement cavity so as to significantly reduce the intensity noise (see SI for details). Spectra collected at different central wavelengths are concatenated to form a total spectrum with broader spectral coverage. All experimental spectra are collected at \SI{50}{Torr} and \SI{20}{\degree C}.

\subsection*{Data processing}
Optical etalon fringes that interfere with the molecular absorption features are removed by notch filtering the experimental data in the post-processing. The zero-absorption baseline is determined by fitting a cubic polynomial to match the upper envelope of the cavity transmitted light spectrum. We note that some molecules (such as methanol) exhibit continuous absorption bands with spectral widths comparable to the spectral bandwidth of the cavity transmitted comb light. In our experiment, these broad molecular absorption components are indistinguishable from the baseline of the light source and are removed during the baseline subtraction process. This results in an alteration of the molecular lineshape, with the extracted absorbance spectrum containing only the sharp molecular absorption features. The experimental absorbance spectrum is fitted using the Beer-Lambert law with the effective interaction length given by $2FL/\pi$, where $L$ is the cavity length and $F$ is the wavelength-dependent cavity finesse (see SI for details on the cavity finesse). The HITRAN database \cite{Gordon2017} and HAPI code \cite{Kochanov2016} are used to obtain the molecular absorption cross sections and generate a simulated spectrum. Starting from an initial guess of the concentrations for the molecules known to be present in the sample, we calculate a simulated absorption spectrum ($A_{sim} = 1 - e^{-\alpha2FL/\pi}$ where $\alpha$ is the total absorption coefficient) containing both the broad and sharp molecular absorption features. The broad absorption feature ($A_{broad}=A_{sim}-A_{sharp}$) is then removed using the same baseline extraction method that is used for processing the experimental spectrum, and the resulting spectrum normalised ($A_{norm} = A_{sharp}/(1-A_{broad})$). The normalised simulated absorbance spectrum is fitted to the normalised experimental absorbance spectrum using a non-linear least squares method. The detection sensitivity analysis is discussed in the SI.

\section*{Acknowledgements}
We thank S. Diddams, M. Thorpe, D. Lesko, B. Bjork, and T. Bui for helpful discussions. This work is funded by the AFOSR (9FA9550-19-1-0148), the NSF (CHE-1665271 and PHY-1734006), and the National Institute of Standards and Technology. J.T. gratefully acknowledges the support of the Lindemann Trust in the form of a Postdoctoral Fellowship.

\section*{Authors contributions}
All authors contributed to the design of the experiment, the discussion of the results, and the writing of the manuscript. J.T., Q.L. and Y.C. built the experimental apparatus, collected the data, and performed the analysis.

\bibliographystyle{ieeetr}
\bibliography{BreathAnalysis}

\begin{thebibliography}{10}

\bibitem{Henderson2018}
B.~Henderson, A.~Khodabakhsh, M.~Metsälä, I.~Ventrillard, F.~M. Schmidt,
  D.~Romanini, G.~A.~D. Ritchie, S.~te~Lintel~Hekkert, R.~Briot, T.~Risby,
  N.~Marczin, F.~J.~M. Harren, and S.~M. Cristescu, ``Laser spectroscopy for
  breath analysis: towards clinical implementation,'' {\em Applied Physics B},
  vol.~124, no.~8, p.~161, 2018.

\bibitem{Lourenco2014}
C.~Lourenço and C.~Turner, ``Breath analysis in disease diagnosis:
  Methodological considerations and applications,'' 2014.

\bibitem{Risby2006}
T.~H. Risby and S.~F. Solga, ``Current status of clinical breath analysis,''
  {\em Applied Physics B}, vol.~85, no.~2, pp.~421--426, 2006.

\bibitem{Modak2011}
A.~S. Modak, ``Barriers to overcome for transition of breath tests from
  research to routine clinical practice,'' {\em Journal of Breath Research},
  vol.~5, no.~3, p.~030202, 2011.

\bibitem{Das2020}
S.~Das and M.~Pal, ``Review--non-invasive monitoring of human health by exhaled
  breath analysis: A comprehensive review,'' {\em Journal of The
  Electrochemical Society}, vol.~167, no.~3, p.~037562, 2020.

\bibitem{Wang2009}
C.~Wang and P.~Sahay, ``Breath analysis using laser spectroscopic techniques:
  Breath biomarkers, spectral fingerprints, and detection limits,'' 2009.

\bibitem{Jones2020}
R.~T. Jones, C.~Guest, S.~W. Lindsay, I.~Kleinschmidt, J.~Bradley, S.~Dewhirst,
  A.~Last, and J.~G. Logan, ``Could bio-detection dogs be used to limit the
  spread of covid-19 by travellers?,'' {\em J Travel Med}, vol.~27, Dec. 2020.

\bibitem{Ruszkiewicz2020}
D.~M. Ruszkiewicz, D.~Sanders, R.~O'Brien, F.~Hempel, M.~J. Reed, A.~C. Riepe,
  K.~Bailie, E.~Brodrick, K.~Darnley, R.~Ellerkmann, O.~Mueller, A.~Skarysz,
  M.~Truss, T.~Wortelmann, S.~Yordanov, C.~L.~P. Thomas, B.~Schaaf, and
  M.~Eddleston, ``Diagnosis of covid-19 by analysis of breath with gas
  chromatography-ion mobility spectrometry - a feasibility study,'' {\em
  EClinicalMedicine}, vol.~29, Dec. 2020.

\bibitem{Shan2020}
B.~Shan, Y.~Y. Broza, W.~Li, Y.~Wang, S.~Wu, Z.~Liu, J.~Wang, S.~Gui, L.~Wang,
  Z.~Zhang, W.~Liu, S.~Zhou, W.~Jin, Q.~Zhang, D.~Hu, L.~Lin, Q.~Zhang, W.~Li,
  J.~Wang, H.~Liu, Y.~Pan, and H.~Haick, ``Multiplexed nanomaterial-based
  sensor array for detection of covid-19 in exhaled breath,'' {\em ACS Nano},
  vol.~14, pp.~12125--12132, Sept. 2020.

\bibitem{Steppert2020}
C.~Steppert, I.~Steppert, W.~Sterlacci, and T.~Bollinger, ``Rapid detection of
  sars-cov-2 infection by multicapillary column coupled ion mobility
  spectrometry (mcc-ims) of breath. a proof of concept study,'' {\em medRxiv},
  p.~2020.06.30.20143347, Jan. 2020.

\bibitem{Smith2014}
D.~Smith, P.~Španěl, J.~Herbig, and J.~Beauchamp, ``Mass spectrometry for
  real-time quantitative breath analysis,'' {\em Journal of Breath Research},
  vol.~8, no.~2, p.~027101, 2014.

\bibitem{Broza2013}
Y.~Y. Broza and H.~Haick, ``Nanomaterial-based sensors for detection of disease
  by volatile organic compounds,'' {\em Nanomedicine}, vol.~8, pp.~785--806,
  May 2013.

\bibitem{Adler2010}
F.~Adler, M.~J. Thorpe, K.~C. Cossel, and J.~Ye, ``Cavity-enhanced direct
  frequency comb spectroscopy: Technology and applications,'' {\em Annual Rev.
  Anal. Chem.}, vol.~3, pp.~175--205, June 2010.

\bibitem{Maslowski2014}
P.~Masłowski, K.~C. Cossel, A.~Foltynowicz, and J.~Ye, ``Cavity-enhanced
  direct frequency comb spectroscopy,'' in {\em Cavity-Enhanced Spectroscopy
  and Sensing} (G.~Gagliardi and H.-P. Loock, eds.), pp.~271--321, Berlin,
  Heidelberg: Springer Berlin Heidelberg, 2014.

\bibitem{Changala2018}
P.~B. Changala, B.~Spaun, D.~Patterson, J.~M. Doyle, and J.~Ye, ``Sensitivity
  and resolution in frequency comb spectroscopy of buffer gas cooled polyatomic
  molecules,'' in {\em Exploring the World with the Laser: Dedicated to Theodor
  Hänsch on his 75th birthday} (D.~Meschede, T.~Udem, and T.~Esslinger, eds.),
  pp.~647--664, Cham: Springer International Publishing, 2018.

\bibitem{Weichman2019}
M.~L. Weichman, P.~B. Changala, J.~Ye, Z.~Chen, M.~Yan, and N.~Picqué,
  ``Broadband molecular spectroscopy with optical frequency combs,'' {\em
  Journal of Molecular Spectroscopy}, vol.~355, pp.~66--78, 2019.

\bibitem{Thorpe2008a}
M.~J. Thorpe and J.~Ye, ``Cavity-enhanced direct frequency comb spectroscopy,''
  {\em Applied Physics B}, vol.~91, no.~3, pp.~397--414, 2008.

\bibitem{Foltynowicz2011}
A.~Foltynowicz, P.~Masłowski, T.~Ban, F.~Adler, K.~C. Cossel, T.~C. Briles,
  and J.~Ye, ``Optical frequency comb spectroscopy,'' {\em Faraday Discuss.},
  vol.~150, no.~0, pp.~23--31, 2011.

\bibitem{Thorpe2008}
M.~J. Thorpe, D.~Balslev-Clausen, M.~S. Kirchner, and J.~Ye, ``Cavity-enhanced
  optical frequency comb spectroscopy: application to human breath analysis,''
  {\em Opt. Express}, vol.~16, no.~4, pp.~2387--2397, 2008.

\bibitem{Ye2004}
J.~Ye and S.~T. Cundiff, eds., {\em Femtosecond Optical Frequency Comb:
  Principle, Operation and Applications}.
\newblock Springer US, 2004.

\bibitem{Schliesser2012}
A.~Schliesser, N.~Picqué, and T.~W. Hänsch, ``Mid-infrared frequency combs,''
  {\em Nature Photonics}, vol.~6, no.~7, pp.~440--449, 2012.

\bibitem{Fortier2019}
T.~Fortier and E.~Baumann, ``20 years of developments in optical frequency comb
  technology and applications,'' {\em Communications Physics}, vol.~2, no.~1,
  p.~153, 2019.

\bibitem{Diddams2020}
S.~A. Diddams, K.~Vahala, and T.~Udem, ``Optical frequency combs: Coherently
  uniting the electromagnetic spectrum,'' {\em Science}, vol.~369, p.~eaay3676,
  July 2020.

\bibitem{Adler2009}
F.~Adler, K.~C. Cossel, M.~J. Thorpe, I.~Hartl, M.~E. Fermann, and J.~Ye,
  ``Phase-stabilized, 1.5 w frequency comb at 2.8-4.8 um,'' {\em Opt. Lett.},
  vol.~34, no.~9, pp.~1330--1332, 2009.

\bibitem{Gordon2017}
I.~E. Gordon, L.~S. Rothman, C.~Hill, R.~V. Kochanov, Y.~Tan, P.~F. Bernath,
  M.~Birk, V.~Boudon, A.~Campargue, K.~V. Chance, B.~J. Drouin, J.-M. Flaud,
  R.~R. Gamache, J.~T. Hodges, D.~Jacquemart, V.~I. Perevalov, A.~Perrin, K.~P.
  Shine, M.-A.~H. Smith, J.~Tennyson, G.~C. Toon, H.~Tran, V.~G. Tyuterev,
  A.~Barbe, A.~G. Császár, V.~M. Devi, T.~Furtenbacher, J.~J. Harrison, J.-M.
  Hartmann, A.~Jolly, T.~J. Johnson, T.~Karman, I.~Kleiner, A.~A. Kyuberis,
  J.~Loos, O.~M. Lyulin, S.~T. Massie, S.~N. Mikhailenko, N.~Moazzen-Ahmadi,
  H.~S.~P. Müller, O.~V. Naumenko, A.~V. Nikitin, O.~L. Polyansky, M.~Rey,
  M.~Rotger, S.~W. Sharpe, K.~Sung, E.~Starikova, S.~A. Tashkun, J.~V. Auwera,
  G.~Wagner, J.~Wilzewski, P.~Wcisło, S.~Yu, and E.~J. Zak, ``The hitran2016
  molecular spectroscopic database,'' {\em Journal of Quantitative Spectroscopy
  and Radiative Transfer}, vol.~203, pp.~3--69, 2017.

\bibitem{Kochanov2016}
R.~V. Kochanov, I.~E. Gordon, L.~S. Rothman, P.~Wcisło, C.~Hill, and J.~S.
  Wilzewski, ``Hitran application programming interface (hapi): A comprehensive
  approach to working with spectroscopic data,'' {\em Journal of Quantitative
  Spectroscopy and Radiative Transfer}, vol.~177, pp.~15--30, 2016.

\bibitem{Lagg1994}
A.~Lagg, J.~Taucher, A.~Hansel, and W.~Lindinger, ``Applications of proton
  transfer reactions to gas analysis,'' {\em International Journal of Mass
  Spectrometry and Ion Processes}, vol.~134, no.~1, pp.~55--66, 1994.

\bibitem{Taucher1995}
J.~Taucher, A.~Lagg, A.~Hansel, W.~Vogel, and W.~Lindinger, ``Methanol in human
  breath,'' {\em Alcoholism: Clinical and Experimental Research}, vol.~19,
  pp.~1147--1150, Oct. 1995.

\bibitem{Adler2010a}
F.~Adler, P.~Masłowski, A.~Foltynowicz, K.~C. Cossel, T.~C. Briles, I.~Hartl,
  and J.~Ye, ``Mid-infrared fourier transform spectroscopy with a broadband
  frequency comb,'' {\em Opt. Express}, vol.~18, no.~21, pp.~21861--21872,
  2010.

\bibitem{Foltynowicz2011a}
A.~Foltynowicz, T.~Ban, P.~Masłowski, F.~Adler, and J.~Ye,
  ``Quantum-noise-limited optical frequency comb spectroscopy,'' {\em PRL},
  vol.~107, p.~233002, Nov. 2011.

\bibitem{Ciaffoni2016}
L.~Ciaffoni, D.~P. O{\textquoteright}Neill, J.~H. Couper, G.~A.~D. Ritchie,
  G.~Hancock, and P.~A. Robbins, ``In-airway molecular flow sensing: A new
  technology for continuous, noninvasive monitoring of oxygen consumption in
  critical care,'' {\em Science Advances}, vol.~2, no.~8, 2016.

\bibitem{Petralia2020}
L.~S. Petralia, A.~Bahl, R.~Peverall, G.~Richmond, J.~H. Couper, G.~Hancock,
  P.~A. Robbins, and G.~A.~D. Ritchie, ``Accurate real-time {FENO} expirograms
  using complementary optical sensors,'' {\em Journal of Breath Research},
  vol.~14, p.~047102, aug 2020.

\bibitem{Maslowski2016}
P.~Maslowski, K.~F. Lee, A.~C. Johansson, A.~Khodabakhsh, G.~Kowzan,
  L.~Rutkowski, A.~A. Mills, C.~Mohr, J.~Jiang, M.~E. Fermann, and
  A.~Foltynowicz, ``Surpassing the path-limited resolution of fourier-transform
  spectrometry with frequency combs,'' {\em PRA}, vol.~93, p.~021802, Feb.
  2016.

\end{thebibliography}


\begin{thebibliography}{1}

\bibitem{Gordon2017}
I.~E. Gordon, L.~S. Rothman, C.~Hill, R.~V. Kochanov, Y.~Tan, P.~F. Bernath,
  M.~Birk, V.~Boudon, A.~Campargue, K.~V. Chance, B.~J. Drouin, J.-M. Flaud,
  R.~R. Gamache, J.~T. Hodges, D.~Jacquemart, V.~I. Perevalov, A.~Perrin, K.~P.
  Shine, M.-A.~H. Smith, J.~Tennyson, G.~C. Toon, H.~Tran, V.~G. Tyuterev,
  A.~Barbe, A.~G. Császár, V.~M. Devi, T.~Furtenbacher, J.~J. Harrison, J.-M.
  Hartmann, A.~Jolly, T.~J. Johnson, T.~Karman, I.~Kleiner, A.~A. Kyuberis,
  J.~Loos, O.~M. Lyulin, S.~T. Massie, S.~N. Mikhailenko, N.~Moazzen-Ahmadi,
  H.~S.~P. Müller, O.~V. Naumenko, A.~V. Nikitin, O.~L. Polyansky, M.~Rey,
  M.~Rotger, S.~W. Sharpe, K.~Sung, E.~Starikova, S.~A. Tashkun, J.~V. Auwera,
  G.~Wagner, J.~Wilzewski, P.~Wcisło, S.~Yu, and E.~J. Zak, ``The hitran2016
  molecular spectroscopic database,'' {\em Journal of Quantitative Spectroscopy
  and Radiative Transfer}, vol.~203, pp.~3--69, 2017.

\bibitem{Kochanov2016}
R.~V. Kochanov, I.~E. Gordon, L.~S. Rothman, P.~Wcisło, C.~Hill, and J.~S.
  Wilzewski, ``Hitran application programming interface (hapi): A comprehensive
  approach to working with spectroscopic data,'' {\em Journal of Quantitative
  Spectroscopy and Radiative Transfer}, vol.~177, pp.~15--30, 2016.

\bibitem{Bevington2003}
P.~Bevington and D.~K. Robinson, {\em Data reduction and error analysis for the
  physical sciences}.
\newblock Boston: McGraw-Hill, 2003.

\bibitem{Foltynowicz2011}
A.~Foltynowicz, T.~Ban, P.~Masłowski, F.~Adler, and J.~Ye,
  ``Quantum-noise-limited optical frequency comb spectroscopy,'' {\em PRL},
  vol.~107, p.~233002, Nov. 2011.

\end{thebibliography}
\end{document}


\title{{\LARGE Supplementary Information for}\\\vspace{3mm}
Ultra-sensitive multi-species spectroscopic breath analysis \\for real-time health monitoring and diagnostics}

\author{Qizhong Liang$^{a,b}$, Ya-Chu Chan$^{a,c}$, P. Bryan Changala$^{d}$, David J. Nesbitt$^{a,b,c}$, Jun Ye$^{a,b,*}$, and Jutta Toscano$^{a,b,}$} \thanks{Corresponding authors: jutta.toscano@jila.colorado.edu, ye@jila.colorado.edu}
\affiliation{$^a$JILA, National Institute of Standards and Technology and University of Colorado, Boulder, CO 80309, USA}
\affiliation{$^b$Department of Physics, University of Colorado, Boulder, CO 80309, USA}
\affiliation{$^c$Department of Chemistry, University of Colorado, Boulder, CO 80309, USA}
\affiliation{$^d$Center for Astrophysics $|$ Harvard \& Smithsonian, Cambridge, MA 02138, USA}
\maketitle

\section*{Cavity-comb frequency stabilization}

Tight locking of the mid-IR comb to the enhancement cavity resonance modes is achieved by actively stabilizing the repetition rate of the comb ($f_{rep}~=~$\SI{136}{MHz}) to half of the enhancement cavity's free spectral range (FSR$~=~$\SI{272}{MHz}), while at the same time locking its carrier-envelope-offset frequency $f_{ceo}$ to a fixed set-point value. For the $f_{rep}$-FSR lock, we generate frequency sidebands by modulating a fast piezo (PZT)-actuated mirror inside the Yb oscillator cavity at the mechanical resonance frequency of \SI{761}{kHz}. Reflected light from the enhancement cavity is spectrally bandpass filtered by a grating and slit to $\approx~$\SI{10}{nm} and measured by a photodetector. The photodetector signal is demodulated at the sideband frequency to generate the Pound-Drever-Hall (PDH) error signal. High frequency error components are fed back to the fast PZT oscillator mirror. In parallel, the low frequency error is fed back onto the enhancement cavity length via a tube piezo (with \SI{1}{kHz} bandwidth). This servo loop ensures that FSR$~=~2\times~f_{rep}$. To stabilize their absolute values, we measure the seventh harmonic of $f_{rep}$ at $7~\times~$\SI{136}{MHz}$~=~$\SI{952}{MHz} with a fast photodetector and beat it against a \SI{1}{GHz} Wenzel quartz oscillator slaved to a \SI{10}{MHz} cesium (Cs) clock. The beat note is phase-locked to an RF signal (produced by a Direct Digital Synthesizer (DDS) referenced to the same \SI{10}{MHz} clock) by feeding back onto the Yb oscillator cavity length. We measure the $f_{ceo}$ by beating the parasitic sum-frequency signal of the pump and the idler generated from the OPO cavity with the pump light spectrally broadened by a supercontinuum fiber. The resulting beat note is again phase-locked with respect to a DDS reference signal by feeding back onto the OPO cavity length. The $f_{ceo}$ lock point frequency is chosen to be far detuned from the critical frequency that gives the highest cavity-comb coupling efficiency. This suppresses the spectral bandwidth of comb light that can be simultaneously coupled into the enhancement cavity. However, mitigation of the frequency-to-amplitude noise conversion results in significantly reduced cavity transmission intensity noise. With all frequency locks engaged, the center wavelength of the cavity-transmitted light can be easily tuned by adjusting the dispersion angle of the grating used for the PDH lock.

\section*{Detection sensitivity analysis}

The detection sensitivity of our experimental setup is characterized with an analysis of the noise equivalent minimum detectable concentration for a series of molecules that are present in human breath. When calculating the detection limit for any specific molecule, we assume that it is the only absorber inside the enhancement cavity. Each individually resolved comb mode gives an independent measurement of the molecular concentration at its respective optical frequency. The measured concentration $N_i$ at the spectral element $i$ is associated with a standard deviation $e_{N_i}~=~\alpha_{min,i}/\sigma_{i}$. Here, $\alpha_{min,i}$ is the minimum detectable absorption (in units of cm$^{-1}$) defined as the standard deviation of the local absorption spectrum measured for an empty cavity, normalized by the effective interaction length $2F_iL/\pi$, where changes in finesse $F_i$ as a function of optical frequency are taken into account (see next section). $\sigma_{i}$ is the molecular absorption cross section (in units of cm$^2$/molecule). For the sensitivity table shown in the main text (Table 1), the cross sectional data from the HITRAN database \cite{Gordon2017,Kochanov2016} is used for C$_2$H$_6$ and CH$_3$OH. The rest of the molecules (eight in total) have their cross sections calculated from the spectral line intensity data with Voigt lineshape profile assumed. The average simulated Lorentzian and Gaussian linewidth components at \SI{293}{K} and \SI{50}{Torr} are \SI{0.31(8)}{GHz} and \SI{0.21(6)}{GHz}, respectively. All cross section spectra were filtered with the same baseline substraction procedure as the measured data, which removes broad, continuous features (see Methods in the main text). By averaging the cross-section spectra locally weighted by $\sigma_i/\alpha_{min,i}$ \cite{Bevington2003}, the total, multi-element detection limit can be shown  to be:

\begin{equation} \label{eq:1}
    N_{min,M>1} = \frac{1}{\sqrt{M}} \Big \langle \Big (\frac{\alpha_{min,i}}{\sigma_i} \Big )^{-2} \Big \rangle_{M} ^{-1/2},
\end{equation} \label{multi-lim}

\noindent
where $\langle ...\rangle_{M}$ denotes the arithmetic mean across $M$ spectral elements, $i = 1,...,M$. The minimum detectable concentration achievable using a single spectral element can be obtained simply by setting $M = 1$. We define the single-element detection limit as the minimum of this value over all spectral elements $i$:

\begin{equation} \label{eq:2}
    N_{min,M=1} = \Big \{ \frac{\alpha_{min,i}}{\sigma_i} \Big \}_{min}.
\end{equation} \label{single-lim}

\noindent
Compared to a measurement performed with a single spectral element, an improvement in the detection sensitivity by a factor of $\sqrt{M}$ can be achieved if a multi-element detection can be performed for $M$ spectral elements giving identical absorption noise and absorption cross section. This result is analogous to signal averaging: through measurements of the same spectral element by $M$ repetitive times the measurement uncertainty can be lowered by a factor of $\sqrt{M}$.

The absorption noise baseline is measured by continuously flowing nitrogen gas through the cavity at a pressure of \SI{50}{Torr}. For this measurement only, a cold trap was used to ensure the cleanest possible cavity. A total of 28 spectra for a total acquisition time of \SI{192}{s} were measured to cover the spectral range of 2810--2945~cm$^{-1}$, over which the minimum detectable absorption is calculated for each spectral element. For the center half of the spectra, where the best lock performance can be achieved due to the high spectral brightness of the cavity incident light, the sensitivity per spectral element (PSE) evaluated for each spectrum yields a statistical mean and standard deviation of 5.4(1.3)~$\times$~10$^{-11}$~cm$^{-1}~$Hz$^{-1/2}$~PSE. Here the sensitivity PSE is calculated from $\alpha_{min}\sqrt{T/M}$, where $\alpha_{min}$ is the minimum detectable absorption in the center part of the spectrum, $T$ is the acquisition time per spectrum and $M$ is the number of spectral elements per spectrum. More details about the calculations of the sensitivity PSE can be found in Ref. \cite{Foltynowicz2011}.

\section*{Cavity finesse measurement}

\begin{figure}
\centering
\includegraphics[width=\textwidth]{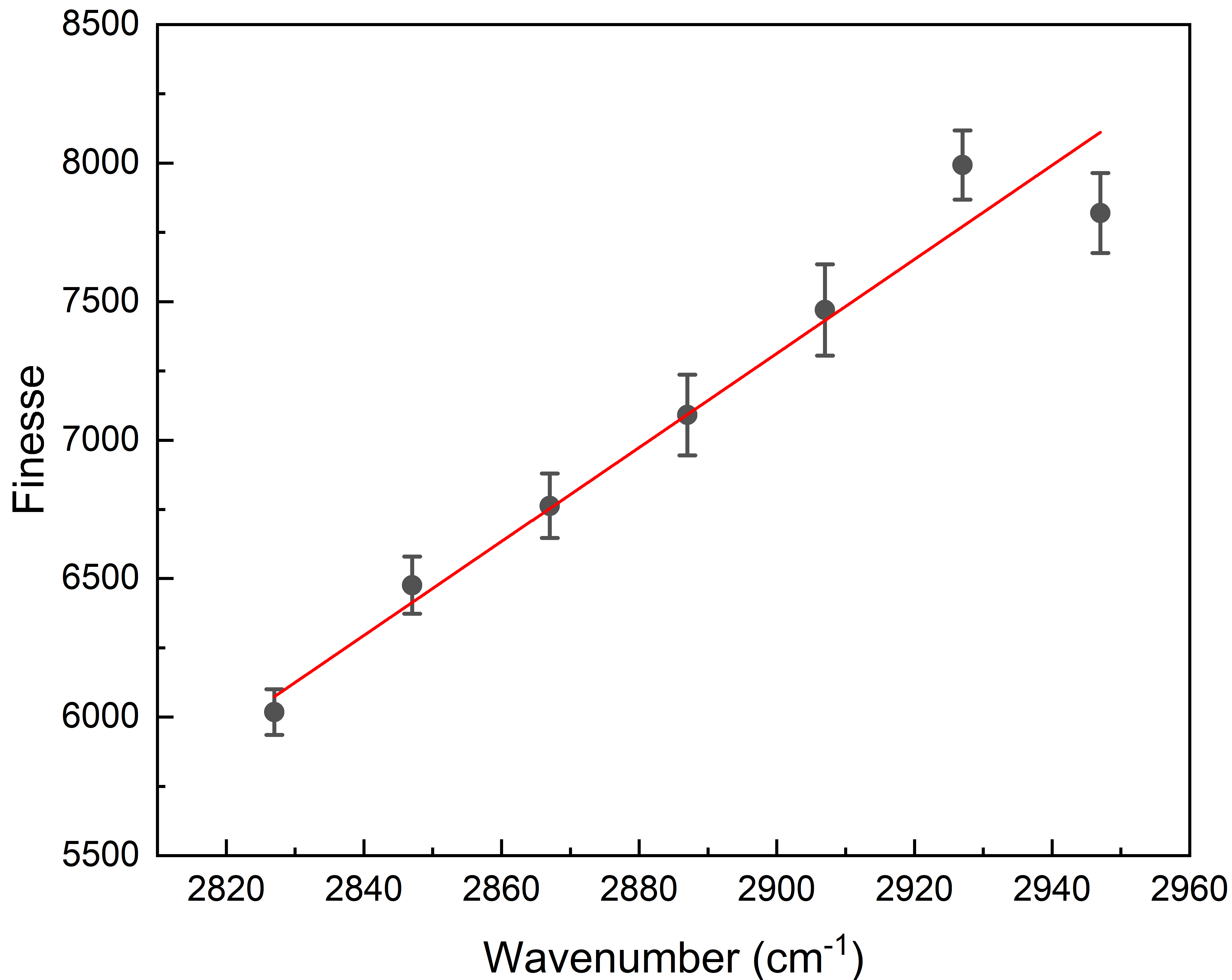}
\caption{Cavity finesse measured at different optical frequencies. Each data point (black) shows the statistical mean and standard error for a total of 20 finesse measurements at each optical frequency. The experimental data is fitted with a linear function (red).}
\label{fig:finesse}
\end{figure}

The cavity finesse is determined from transmission ring down measurements performed by rapidly sweeping the comb light through the cavity resonances (Fig. \ref{fig:finesse}). To avoid issues related to cavity mirror dispersion, the cavity transmission light is spectrally dispersed by a grating to measure the finesse for one narrow spectral region at a time. This process is repeated at different grating angles to measure the finesse as a function of optical wavelength. A linear fit of the measured finesse curve is used for the molecular line fitting (see main text) and for the detection sensitivity analysis (see previous section).

\bibliographystyle{ieeetr}
\bibliography{BreathAnalysisSUPPL}